# A discussion on the critical electric Rayleigh number for AC electrokinetic flow of binary fluids in divergent microchannel

Jin'an Pang(庞晋安), Yu Han(韩煜), Bo Sun(孙博), Wei Zhao(赵伟) †

State Key Laboratory of Photon-Technology in Western China Energy, International Collaborative Center on Photoelectric Technology and Nano Functional Materials, Laboratory of Optoelectronic Technology of Shaanxi Province, Institute of Photonics & Photon Technology, Northwest University, Xi'an 710127, China

**ABSTRACT**

Electrokinetic (EK) flow is a type of flow driven or manipulated by electric body forces, influenced by various factors such as electric field intensity, electric field form, frequency, electric permittivity/conductivity, fluid viscosity and etc. The diversity of dimensionless control parameters, such as the electric Rayleigh number, complicates the comparison of EK flow stability. Consequently, comparing the performance and cost of micromixers or reactors based on EK flow is challenging, posing an obstacle to their industrial and engineering applications. In this investigation, we theoretically derived a new electric Rayleigh number ($Ra_e$) that quantifies the relationship among electric body forces, fluid viscosity, and ion diffusivity, based on a tanh model of electric conductivity distribution. The calculation results indicate that the new $Ra_e$ exhibits richer variation with the control parameters and better consistency with previous experimental reports. We further conducted experimental studies on the critical electric Rayleigh number ($Ra_{ec}$) of AC EK flow of binary fluids in a divergent microchannel. The experimental variations align well with the theoretical predictions, particularly the existence of an optimal AC frequency and electric conductivity ratio, demonstrating that the tanh model can better elucidate the underlying physics of EK flow. With the new electric Rayleigh number, we found that EK flow in the designed divergent microchannel has a much smaller $Ra_{ec}$ than previously reported, indicating that EK flow is more unstable and thus more suitable for applications in micromixers or reactors in industry and engineering.

## I. INTRODUCTION

Microfluidic device is a pivotal member of lab on a chip (LOC), facilitating the manipulation of fluids ranging from microliters to picoliters[1] in fields such as biomedical[2-4] and chemical engineering,[5-8] including applications like polymerase chain reaction (PCR)[9], synthesis of nanoparticles[10] and DNA sequencing chips.[11]

Micromixer represents a significant category of microfluidic devices. According to the mechanism of mixing enhancement, micromixers are typically classified into two types: passive and active[12]. Passive micromixers enhance mixing through the folding and stretching of fluids within specially designed microchannel geometries, requiring no additional driving mechanisms beyond a pressure gradient to induce flow disturbance.[13-16] While passive micromixers often have specific functions, they generally lack versatility and exhibit relatively low mixing efficiency. In 2002, Stroock et al.[17] introduced a staggered herringbone mixer, discovering that at a Péclet number of $2 \times 10^3$, the length required for 90% mixing was 7 mm, with the channel length needed for mixing increasing logarithmically with the Péclet number.

Active micromixers can significantly enhance mixing by applying external forces, such as electrostatic force,[18] Lorentz force[19] and acoustic pressure.[20] Electrokinetic (EK) micromixer is a specific type of active micromixer that enhances mixing by EK flow driven by electrostatic force, and experiences a rapid advancement since 1990s.[21] In 1998, Baygents et al.[22] investigated the neutral stability curves for EK flow under a DC electric field that parallel to the electric conductivity interface. In their investigation, the status of EK flow is characterized by an electric Rayleigh number ($Ra_e$, see Table I), which is determined by the electric intensity, the interface width and the gradient of electric conductivity. In

**TABLE I**. Definitions of $Ra_e$ in different EK flow model. The red arrow indicates the flow direction, while the green arrow indicates the direction of the electric field.

| Researchers | Baygents et al (1998) [22] | Chen et al (2005)[23] | Posner et al (2006)[24] | Wang et al (2014)[25] |
|---|---|---|---|---|
| $Ra_e$ | $\dfrac{\varepsilon E^2}{D\mu}d^2\dfrac{\Delta\sigma}{\sigma_1}$ | $\dfrac{\varepsilon E^2}{4D\mu}\dfrac{dh^2}{\Delta y}\dfrac{(\lambda-1)^2}{(\lambda+1)^2}$ | $\dfrac{\varepsilon E^2}{D\mu}h^2\dfrac{\lambda-1}{\lambda}\nabla^*\sigma^*\Big|_{ma}$ | $\dfrac{\varepsilon E^2}{D\mu}w^2\dfrac{\Delta\sigma}{\sigma_1}$ |
| Diagram of the EK flow | | | | |

| Researchers | Wu et al (2015)[26] | Yoshikawa et al (2015) [27] | Hassen et al (2020)[28] | Nan et al (2022) [21] |
|---|---|---|---|---|
| $Ra_e$ | $\dfrac{\varepsilon E d}{K\mu}$ | $\dfrac{e\Delta\theta l^3}{\mu\kappa}\nabla\dfrac{\varepsilon_{ref}E^2}{2}$ | $\dfrac{\varepsilon E d}{K\mu}$ | $\dfrac{4\varepsilon E^2}{D\mu}w^2\dfrac{\lambda-1}{\lambda+1}(1-\beta^2)$ |
| Diagram of the EK flow | | | | |

$E$: Applied electric field; $d$: Channel width; $\sigma_1$: Lower electric conductivity; $\sigma_2$: Higher electric conductivity; $\Delta\sigma = \sigma_2 - \sigma_1$; $D$: Effective diffusivity; $\varepsilon$: Electric permittivity; $\mu$: Dynamic viscosity; $\kappa$: Thermal diffusivity; $\Delta y$: Interface width; $h$: Channel height; $\lambda = \sigma_2/\sigma_1$: Electric conductivity ratio; $K$: Ionic mobility; $b$: Initial interface width; $w$: Initial width at the inlet; $l$: Characteristic length scale of the flow; $\theta$: Temperature deviation; $\varepsilon_{ref}$: Electric permittivity at reference temperature; $e$: Coefficient; $\beta = 2\omega\varepsilon/(\sigma_1+\sigma_0)$: Dimensionless number to evaluate the influence of frequency; $\nabla^*\sigma^*|_{max} \approx 1 - e^{-\left(\frac{b}{\Delta y}\right)^2} - e^{-\left(\frac{d-b}{\Delta y}\right)^2}$: Normalized maximum conductivity gradient;

2001, Oddy et al.[29] designed a delicate micromixer to promote mixing through EK instability induced by sinusoidal electroosmotic flow. Subsequently, Lin et al.[30] showed a plot of the growth rates of the most unstable eigenfunction in the wavenumber ($k$)—$Ra_e$ space. They found the critical electric field required for mixing decreases as the electric conductivity ratio increases. In 2005, Chen et al.[23] defined a $Ra_e$ (refer to Table I) for DC EK flow with electric conductivity difference in a T-shaped microchannel with a high aspect ratio. By experimental and analytical studies, they found the presence of both convective and absolute instabilities of EK flow. When the applied $Ra_e$ is beyond a certain critical threshold, say $Ra_{ec} \approx 10$, the flow becomes convective unstable where the disturbance only grows downstream. When $Ra_e$ is beyond a second $Ra_{ec} \approx 160$, the flow becomes absolute unstable, with the disturbances propagating both upstream and downstream simultaneously. In 2006, Posner et al.[31] studied the EK instability in a cross-shaped microchannel flow diagram. By defining a modified local electric Rayleigh number, characterized by the charge density as a function of the maximum electric conductivity gradients, they determined that the critical electric Rayleigh number $Ra_{ec} = 205$.

In 2014, Wang et al. designed an elaborate Y-type EK micromixer[5, 25, 32] and demonstrated that micro electrokinetic (μEK) turbulence could be generated under an alternating-current (AC) electric field with 100 kHz frequency. However, the additional control parameter, e.g. AC frequency, makes the comparison of working conditions with the previous reports unattainable. Further investigations by Nan et al.[21] demonstrate that the conditions of inducing the most unstable mode of EK flow is inconsistent to that of generating highly-developed EK turbulence. This further indicates the difficulty to advance a universal control parameter for EK turbulence, e.g. electric Rayleigh number ($Ra_e$). Although the $Ra_e$ deduced from the Zhao-Wang model[33] can evaluate the flow status in a highly-developed AC EK turbulence, they are not appropriate indicators to show how the EK flow evolves from its initial status, which could be more crucial for engineering applications.

The strong diversity of dimensionless control parameters, e.g. $Ra_e$, could be attributed to the broad factors that influencing the electric body force in EK flow, including the form, intensity and frequency of electric field, the geometry of the flow field, the distribution of electric conductivity and permittivity, pH value, electrode property and etc. The entangled dynamics and interfacial phenomenon make the problem more complicated to reach a common sense on the format of $Ra_e$.

In this investigation, a novel $Ra_e$ that describing the transition of AC EK flow of binary fluids from stable to unstable states is theoretically analyzed and experimentally evaluated in a divergent microchannel. We focus on how $Ra_e$ is influenced by the physical and chemical conditions, e.g. electric field intensity, AC frequency, electric conductivity ratio and interface width. A theoretical analysis on electric Rayleigh number is conduct first. Then, the theory is adopted to predict the critical electric Rayleigh number $Ra_{ec}$ of AC EK flow in a divergent microchannel. This study aims to show the newly developed electric Rayleigh number can better elucidate the underlying physics of how the electric body force perturbs the interface of the binary fluids. With the electric Rayleigh number, people can more precisely predict $Ra_e$, facilitating the comparison of physical and chemical conditions of generating EK flow. This is important for enhancing our understanding of the mechanisms of EK turbulence and promoteing its application in microfluidics.

## II. Theory and dimensionless parameters

### A. Control equations

The EK flow can be described by Navier-Stokes equations as shown below[34, 35]

$$\rho\left(\frac{\partial}{\partial t}\vec{u} + \vec{u}\cdot\nabla\vec{u}\right) = -\nabla p + \vec{F_e} + \mu\nabla^2\vec{u} \quad (1)$$

$$\vec{F_e} = \rho_e\vec{E} - \frac{1}{2}(\vec{E}\cdot\vec{E})\nabla\varepsilon + \frac{1}{2}\nabla\left[\rho\vec{E}\cdot\vec{E}\left(\frac{\partial\varepsilon}{\partial\rho}\right)_T\right] \quad (2)$$

$$\rho_e = \nabla\cdot\varepsilon\vec{E} \quad (3)$$

$$\frac{\partial}{\partial t}\sigma + \vec{u}\cdot\nabla\sigma = D_\sigma\nabla^2\sigma \quad (4)$$

where $\rho$ is fluid density, $\mu$ is dynamic viscosity, $\vec{F_e}$ is electric body force (EBF). $\varepsilon$ and $\sigma$ are electric permittivity and electric conductivity of the solution respectively, while $\rho_e$ is the charge density. The instantaneous velocity vector, $\vec{u} = u\vec{i} + v\vec{j} + w\vec{k}$, consist of $u$, $v$, $w$ as the velocity components in streamwise, spanwise and vertical directions respectively, where $\vec{i}$, $\vec{j}$ and $\vec{k}$ are the unit vectors in these corresponding directions. $D_\sigma$ stands for the diffusion coefficient. Furthermore, we also assume the fluid to be incompressible, i.e. $\nabla\cdot\vec{u} = 0$.

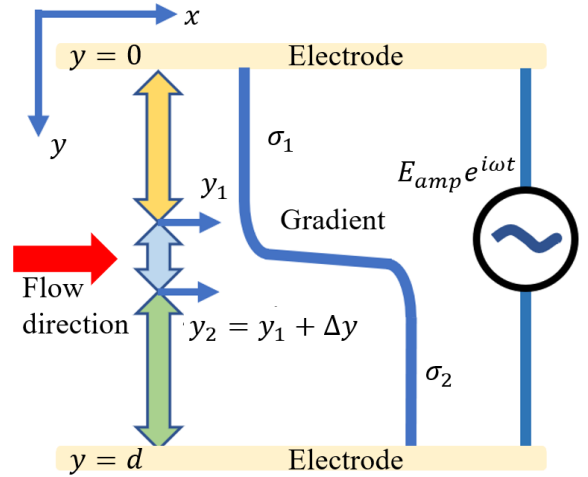

**FIG. 1.** Schematic diagram of electric conductivity distribution along y direction in initial stage. $y_1$ denotes the width with electric conductivity $\sigma_1$. $\Delta y$ represents the interface width with electric conductivity gradient. $d - y_2$ is the width with electric conductivity $\sigma_2$. Note that, given the flow rate is the same, here $y_1 = d - y_2$.

In this model, we consider that the electric field is perpendicular to the interface of the binary fluids, as diagramed in Fig. 1, with the AC frequency ($f_f$) much higher than the response frequency ($f_r$) of the flow.[36] The fluids share identical viscosity, temperature and electric permittivity, yet differ in electric conductivity, denoted as $\sigma_1$ and $\sigma_2$ ($\sigma_2 \geq \sigma_1$). Since the fluid is incompressible, $\frac{1}{2}\nabla\left[\rho\vec{E}\cdot\vec{E}\left(\frac{\partial\varepsilon}{\partial\rho}\right)_T\right] = 0$. Accordingly, Eq. (2) becomes

$$\vec{F_e} = \rho_e\vec{E} - \frac{1}{2}(\vec{E}\cdot\vec{E})\nabla\varepsilon \quad (5)$$

When an electric field is applied, initially, the interface between the binary fluids is slightly disturbed by the EBF. The ultrasmall flow disturbance has three consequences. First, the distribution of electric conductivity has negligible difference from the case without EBF. Second, the convective transport of charge is much weaker than that of the electric field. Therefore, the charge conservation equation[37] becomes

$$\frac{\partial \rho_e}{\partial t} + \nabla \cdot \sigma \vec{E} = 0 \qquad (6)$$

The third is that the disturbance of electric conductivity is negligible as well. Consequently, the electric field can be approximated as a 1D model in the y-direction, $E_y$. Thus, Eq. (6) can be rewritten in the following as

$$\frac{\partial}{\partial y} \sigma^* E_y = 0 \qquad (7)$$

where $\sigma^* = i\omega\varepsilon + \sigma$ is the complex electric conductivity, with $\omega = 2\pi f_f$ denoting the angular frequency. So considering the complex Ohmic law, $\sigma^* E_y = J^*$, it is constant for $y$ where $J^*$ is complex electric current density.

**B. Electric body force**

In EK flow, EBF is the most important quantities that determines the state of flow, however, influenced by various factors. At the initial stage, the time-averaged AC EBF per unit volume can be rewritten as[34]

$$\vec{F}_{e,ac} = \frac{1}{2}\mathcal{R}e\left[(\nabla\varepsilon \cdot \vec{E} + \varepsilon\nabla \cdot \vec{E}) \cdot \vec{\tilde{E}} - \frac{1}{2}E^2\nabla\varepsilon\right]$$

where $\vec{\tilde{E}}$ is the complex conjugation of electric field intensity and $\mathcal{R}e$ is the real part. We are unable to use a local $\vec{F}_{e,ac}$ to evaluate the influence of EBF on EK flow. To characterize the influence of EBF, especially aiding to engineering applications, a spatially averaged EBF [36] on the interface with a width of $\Delta y$ is applied as

$$F_{e,\Delta y} = \langle F_{e,ac}\rangle_{\Delta y} = \frac{\varepsilon}{4\Delta y}\frac{|J^*|^2}{|\sigma^*|^2}\bigg|_{y_1}^{y_1+\Delta y} \qquad (8)$$

where $\langle \cdot \rangle_{\Delta y}$ indicates the spatial averaging on $\Delta y$ in the $y$ direction. The complex current density is conserved in space as

$$J^* = \sigma^*_{\Delta y} E_{y,\Delta y} = \sigma^*_d E_{y,d} \qquad (9)$$

where $\sigma^*_{\Delta y} = 1/\langle 1/\sigma^*\rangle_{\Delta y}$ and $E_{y,\Delta y} = \langle E_y\rangle_{\Delta y}$ are the complex electric conductivity and electric field intensity averaged on the interface with $\Delta y$ respectively. Similarly, $\sigma^*_d = 1/\langle 1/\sigma^*\rangle_d$ and $E_{y,d} = E_{amp}e^{i\omega t}$ are the complex electric conductivity and electric field intensity averaged across the channel width $d$ respectively, where $E_{amp} = V_{amp}/d$ is the amplitude of the AC electric field and $V_{amp}$ is the voltage amplitude between the two electrodes as shown in the Fig. 1. Thus, $|J^*|^2 = J^* \cdot J = \sigma_d \sigma_d^* E_{y,d} E_{y,d}^* = \sigma_d \sigma_d^* E_{amp}^2$. $\sigma_d^*$ is strictly dependent on $\omega$, $\varepsilon$ and the spatial distribution of $\sigma$.

Here, we consider an initial case of the EK flow, where the EBF is too weak to induce a perturbation of $\sigma$. Thus, the distribution of $\sigma$ can be approximated to be steady. Furthermore, we utilize a tanh distribution (Fig. 1) of electric conductivity to demonstrate how the EBF and the corresponding dimensionless parameters is influenced by the physical quantities. If the electric conductivity has a tanh distribution as below

$$\sigma(y) = \sigma_1 + \frac{e^{(2y-d)/y_s}}{e^{(2y-d)/y_s} + 1}(\sigma_2 - \sigma_1) \qquad (10)$$

where $y_s$ is a shape factor to characterize the distribution of $\sigma$. In the tanh model, to stimulate an interface width of $\Delta y$, approximately we have $y_s \approx \Delta y/10$. According to Eq. (10), $\sigma_d^*$ can be calculated as

$$\sigma_d^* = \frac{1}{\left\langle\frac{1}{\sigma^*}\right\rangle_d} = \frac{2d\sigma_1^*\sigma_2^*}{y_s(\sigma_1 - \sigma_2)\ln\frac{\sigma_1^* + \sigma_2^* e^{d/y_s}}{\sigma_1^* + \sigma_2^* e^{-d/y_s}} + 2d\sigma_2^*} \qquad (11)$$

Furthermore, according to $\sigma^* = \sigma + i\omega\varepsilon$ and Eq. (10), we have

$$\frac{1}{|\sigma^*|^2}\bigg|_{y_1}^{y_1+\Delta y} = 16\frac{(1-\lambda^2)}{\sigma_1^2}\frac{B_{tanh}}{A_{tanh}} \qquad (12)$$

where $\lambda = \sigma_2/\sigma_1$ is electric conductivity ratio, $\chi = 2\omega\varepsilon/(\sigma_1 + \sigma_2)$ is a dimensionless frequency, $m = \chi(1+\lambda)$, $A_{tanh} = [4(e^{\Delta y/y_s}\lambda + 1)^2 + m^2(e^{\Delta y/y_s} + 1)^2][4(e^{\Delta y/y_s} + \lambda)^2 + m^2(e^{\Delta y/y_s} + 1)^2]$ and $B_{tanh} = (e^{2\Delta y/y_s} - 1)(e^{\Delta y/y_s} + 1)^2$. By substituting Eq. (12) into (8), with binomial expansion, we have

$$F_{e,\Delta y} = \frac{\varepsilon E_{amp}^2}{\Delta y}\frac{B_{tanh}(1-\lambda^2)(4\lambda^2 + m^2)(4 + m^2)}{A_{tanh}C_{tanh}} \qquad (13)$$

where $C_{tanh} = (1-\lambda)^2(y_s/d)^2 q_{tanh} p_{tanh} + (y_s/d)(1-\lambda)(2\lambda - im)q_{tanh} + (y_s/d)(1-\lambda)(2\lambda + im)p_{tanh} + 4\lambda^2 m^2$, with $q_{tanh} = \ln\frac{2+im+(2\lambda+im)e^{d/y_s}}{2+im+(2\lambda+im)e^{-d/y_s}}$ and $p_{tanh} = \ln\frac{2-im+(2\lambda-im)e^{d/y_s}}{2-im+(2\lambda-im)e^{-d/y_s}}$.

**TABLE II.** Comparison between the linear model and the tanh model.

| | Linear model | Tanh model |
|---|---|---|
| Schematic diagram of electric conductivity distribution | | |
| Electric conductivity distribution $\sigma$ | $\begin{cases} \sigma_1, (0 \leq y < y_1) \\ \dfrac{\sigma_2 - \sigma_1}{\Delta y}(y - y_1) + \sigma_1, (y_1 \leq y < y_2) \\ \sigma_2, (y_2 \leq y \leq d) \end{cases}$ | $\sigma_1 + \dfrac{e^{(2y-d)/y_s}}{e^{(2y-d)/y_s} + 1}(\sigma_2 - \sigma_1)$ |
| $F_{e,\Delta y}$ | $\dfrac{4\varepsilon E_{amp}^2}{\Delta y}\dfrac{(1-\lambda)^2(1-\lambda^2)}{A_{linear}}$ | $\dfrac{\varepsilon E_{amp}^2}{\Delta y}\dfrac{B_{tanh}(1-\lambda^2)(4\lambda^2+m^2)(4+m^2)}{A_{tanh}C_{tanh}}$ |
| $Ra_e$ | $\dfrac{4\varepsilon E_{amp}^2 \Delta y^2}{\rho\nu D_e}\dfrac{(1-\lambda)^2(1-\lambda^2)}{A_{linear}}$ | $\dfrac{\varepsilon E_{amp}^2 \Delta y^2}{\rho\nu D_e}\dfrac{B_{tanh}(1-\lambda^2)(4\lambda^2+m^2)(4+m^2)}{A_{tanh}C_{tanh}}$ |

$A_{linear} = 4(1-\Delta y/d)^2(\lambda-1)^2(\lambda^2+2\lambda+1+m^2)$
$\quad + 2\Delta y/d\,(1-\Delta y/d)(\lambda-1)[2(q_{linear}+p_{linear})$
$\quad + im(q_{linear}-p_{linear})](m^2+2+2\lambda^2)$
$\quad + (\Delta y/d)^2 q_{linear}p_{linear}(16\lambda^2+4m^2+4m^2\lambda^2+m^4)$

$q_{linear} = \ln\dfrac{2\lambda+im}{2+im}$

$p_{linear} = \ln\dfrac{2\lambda-im}{2-im}$

$A_{tanh} = [4(e^{\Delta y/y_s}\lambda+1)^2 + m^2(e^{\Delta y/y_s}+1)^2][4(e^{\Delta y/y_s}+\lambda)^2 + m^2(e^{\Delta y/y_s}+1)^2]$

$B_{tanh} = (e^{2\Delta y/y_s}-1)(e^{\Delta y/y_s}+1)^2$

$C_{tanh} = (1-\lambda)^2(y_s/d)^2 q_{tanh}p_{tanh}$
$\quad + (y_s/d)(1-\lambda)(2\lambda-im)\,q_{tanh}$
$\quad + (y_s/d)(1-\lambda)(2\lambda+im)p_{tanh}$
$\quad + 4\lambda^2 m^2$

$q_{tanh} = \ln\dfrac{2+im+(2\lambda+im)e^{d/y_s}}{2+im+(2\lambda+im)e^{-d/y_s}}$

$p_{tanh} = \ln\dfrac{2-im+(2\lambda-im)e^{d/y_s}}{2-im+(2\lambda-im)e^{-d/y_s}}$

### C. Electric Grashof number and electric Rayleigh number

The electric Grashof number and the electric Rayleigh number are two commonly used dimensionless parameters to describe the state of EK flow. The former represents the ratio of EBF relative to viscous force, while the latter further considers the influence of ion diffusion. According to Eq. (13), the electric Grashof number[38] ($Gr_e$) can be quantified by

$$Gr_e = \dfrac{F_{e,\Delta y}\Delta y^3}{\rho\nu^2}$$

$$= \dfrac{\varepsilon E_{amp}^2 \Delta y^2}{\rho\nu^2}\dfrac{B_{tanh}(1-\lambda^2)(4\lambda^2+m^2)(4+m^2)}{A_{tanh}C_{tanh}} \quad (14)$$

The electric Rayleigh number ($Ra_e$) is further obtained according to $Ra_e = Gr_e Sc$, where $Sc = \nu/D_\sigma$ is Schmidt number. Then, we have

$$Ra_e = \frac{\varepsilon E_{amp}^2 \Delta y^2}{\rho \nu D_e} \frac{B_{tanh}(1-\lambda^2)(4\lambda^2+m^2)(4+m^2)}{A_{tanh}C_{tanh}} \quad (15)$$

$Ra_e$ in Eq. (15) is evaluated from the overall distribution of electric conductivity. It complexly interplays with $\lambda$, $\chi$ and the initial distribution (through $y_1$, $y_2$ and $\Delta y$) of electric conductivity, in a relatively nonintuitive manner. This is apparently distant from the previous investigations (see Table I), where the electric Rayleigh number are simply evaluated through either simple local quantities or a linear distribution of control scalars. For comparison, a linear model for electric conductivity distribution is also advanced in this investigation, as summarized in Table. II.

## II. Numerical calculation of $Ra_e$ in both linear and tanh model

In this section, as indicated by $Ra_e$ in the Table. II, we demonstrate how the following parameters, e.g. $E_{amp}$, $\Delta y$, $\lambda$ and $\chi$, contribute to the variations of $Ra_e$. Both the linear mode and tanh model are theoretically compared to reveal their difference on calculating $Ra_e$. To be consistent to experiments, some relevant parameters are given as: $d = 650$ μm, $\varepsilon$ is $7.1 \times 10^{-10}$ F/m, $\rho$ is $10^3$ kg/m$^3$, $\nu$ is $10^{-6}$ m$^2$/s, $\sigma_1 + \sigma_2$ is $2.395 \times 10^{-1}$ S/m and $D_\sigma$ is $5 \times 10^{-10}$ m$^2$/s.

### A. Electric conductivity ratio

First, the influence of $\lambda$ on $Ra_e$ in the initial stage of EK flow is analyzed, as shown in Fig. 2. Under the linear model, as depicted in Fig. 2(a), $Ra_e$ increases rapidly and monotonically with increasing $\lambda$, eventually reaching a saturation point at $\lambda = O(10^2)$. This indicates a higher probability of reaching an unstable state at higher $\lambda$, accordingly, easier to get into a transition. There is no optimal $\lambda$ related to the highest $Ra_e$ in the linear model.

However, under the tanh model, $Ra_e$ first increases with $\lambda$ and then decreases at relatively smaller $\chi$, as depicted in Fig. 2(b). This model provides us with the possibility of identifying an optimal $\lambda$ to reach the highest $Ra_e$, for effectively perturbing flow and improving momentum and scalar transport. Relative to the linear model without optimal $\lambda$, the tanh model that predicts optimal $\lambda$ can provide better consistency with previous experiments. For instance, Wang et al[38] found when the applied AC frequency was 100 kHz, i.e.

$\chi = 3.7 \times 10^{-3}$, an ultrafast EK mixing can be achieved with an optimal $\lambda = 5000$. Nan et al[21] also observed different optimal $\lambda$ for perturbating the electric conductivity interface under different $\chi$.

From Fig. 2(b), it should be noted that only when $\chi \geq 3.7 \times 10^{-3}$ (i.e. $f_f = 100$ kHz), $Ra_e$ increases monotonically with $\lambda$. To avoid the violation of $\tau_{rex} \ll 1/\omega$ (where $\tau_{rex} = 1.7 \times 10^{-5}$ is the charge relaxation time)[36], a larger $\chi$ exceeding $3.7 \times 10^{-2}$ is not applied in this study. In the range $1.1 \times 10^{-6} \ll \chi \ll 3.7 \times 10^{-2}$, we surprisingly find a special value of $\lambda_c \approx 200$ by the tanh model. When $\lambda \leq \lambda_c$, $Ra_e$ is insensitive to $\chi$.

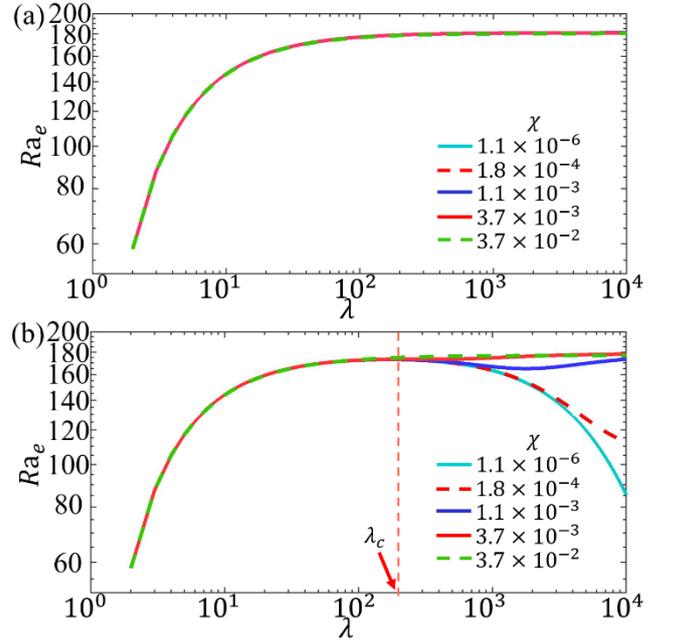

**FIG. 2.** $Ra_e$ calculated from both linear and tanh models under different $\chi$, where $d = 650$ μm, $\Delta y = \sqrt{4D_\sigma t} = 11.4$ μm and $E_{amp} = 3.1 \times 10^4$ V/m, where $t = x/U_e$ is the diffusion time of the interface. Here, $U_e$ is the mean flow velocity at the centerline of the microchannel. $x = 0.2d$ is the measured position. All these data are consistent to the experimental conditions in the following. (a) $Ra_e$ calculated by the linear model vs $\lambda$. (b) $Ra_e$ calculated by the tanh model vs $\lambda$.

### B. AC frequency

The insensitivity of $Ra_e$ to $\chi$ at $\lambda \leq \lambda_c$ can be more visible in Fig. 3(b). All the plots at $\lambda \leq \lambda_c$ are exactly flat lines in the wide range of $\chi$. However, when $\lambda > \lambda_c$, e.g. 2000 and 5400, $Ra_e$ changes significantly with $\chi$. It stays almost unchanged with $\chi$ first, then rapidly increases until reaching another plateau.

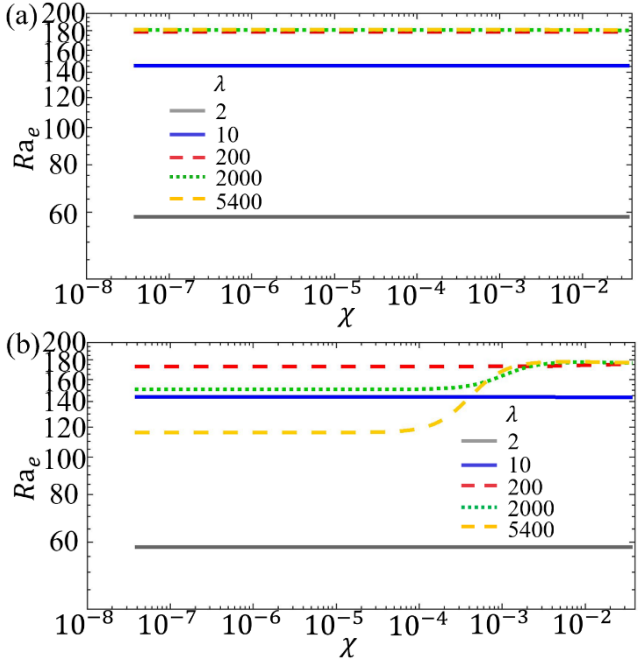

**FIG. 3.** $Ra_e$ calculated from both linear and tanh models under different $\lambda$, where $d = 650$ μm, $\Delta y = 11.4$ μm and $E_{amp} = 3.1 \times 10^4$ V/m. (a) $Ra_e$ calculated by the linear model vs $\chi$. (b) $Ra_e$ calculated by the tanh model vs $\chi$.

In contrast, the variation of $Ra_e$ with $\chi$ is not observed in Fig. 3(a), indicating the linear model has no capability to reveal the influence of $\chi$ (or AC frequency) on $Ra_e$. This accounts for why in previous experiments[21], even though ultrafast EK mixing can be observed at $\chi > 3.7 \times 10^{-4}$ (or $f_f > 10$ kHz), we still cannot get an apparently larger $Ra_e$ through the linear model.

### C. Interface width

As illustrated by a series of investigations[21, 22, 38], a larger gradient of electric conductivity is crucial for generating a larger EBF, accordingly a smaller $\Delta y$ is preferred. Thus, a smaller $\Delta y$ is corresponding to a larger $Ra_e$. For this purpose, we theoretically studied the effect of $\Delta y$ on $Ra_e$. As depicted in Fig. 4, both models predict a larger $Ra_e$ with increasing $\Delta y$ as anticipated. The distinction lies in the fact that the tanh model can better distinguish the influence of $\Delta y$, as can be observed in Fig. 4(a, b).

### D. Electric field intensity

In contrast to $\lambda$, $\chi$ and $\Delta y$ which have complex relationship with $Ra_e$, the quadratic effect of electric field intensity is quite straightforward, i.e. $Ra_e \sim E_{amp}^2$ according to Eq. (15). This is also observable from Fig. 5(a, b). The difference between the linear and tanh model is that the latter is distinguishable for different $\chi$, while the former is not.

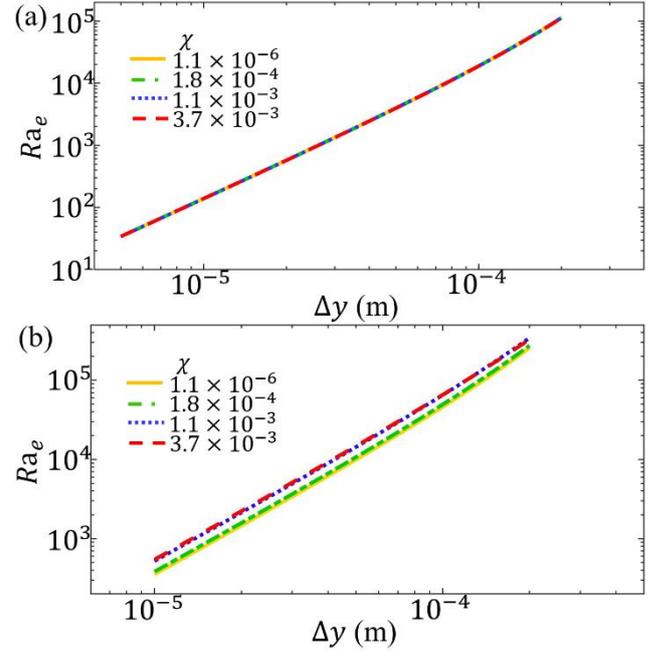

**FIG. 4.** $Ra_e$ calculated from both linear and tanh models under different $\chi$, where $d = 650$ μm, $\lambda = 5400$ and $E_{amp} = 3.1 \times 10^4$ V/m. (a) $Ra_e$ calculated by the linear model vs $\Delta y$. (b) $Ra_e$ calculated by the tanh model vs $\Delta y$.

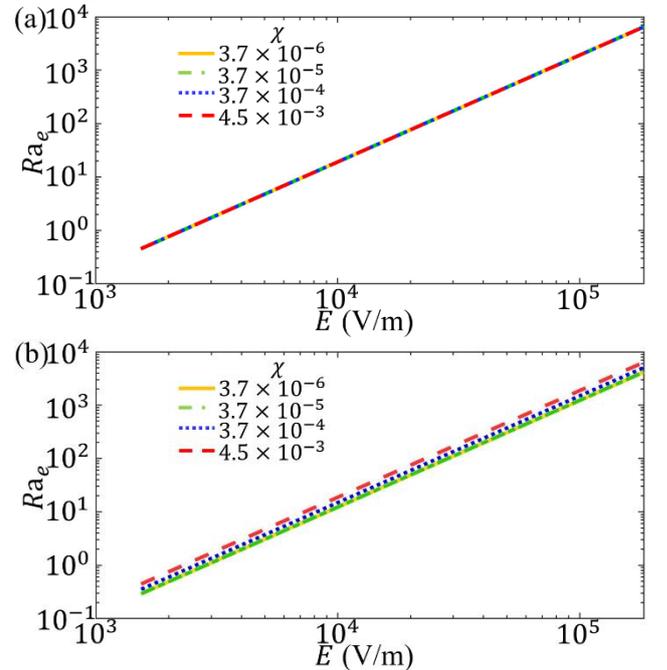

**FIG. 5.** $Ra_e$ calculated from both linear and tanh models under different $\Delta y$, where $d = 650$ μm, $\chi = 3.7 \times 10^{-3}$ and $\lambda = 5400$. (a) $Ra_e$ calculated by the linear model vs $E$. (b) $Ra_e$ calculated by the tanh model vs $E$.

## III. Experimental setup and results

### A. Experimental setup

In this section, relying on the electric Rayleigh number $Ra_e$ under the tanh model, we experimentally investigate the conditions when EK flow becomes unstable under AC electric field. To this end, the EK flow velocity is measured by laser induced fluorescence photobleaching anemometer (LIFPA), which has ultra-high spatial (~203 nm) and temporal (~4 μs) resolution, as diagramed in Fig. 6(a). The system uses a 405nm continuous wave (CW) laser, with a laser power of 11 mW, as light source. An acousto-optic modulator (AOM) is applied to realize temporal control of the beam. The beam in turn passes through a special light filter (SLF) and a dichroic mirror (DM), and is then reflected into an objective lens (OL) to excite the fluorescent solution in the EK flow microchip. The fluorescent solution is Coumarin 102. The fluorescence emitted by the solution is filtered out by two bandpass filters (BP, 470/100 and 470/10 nm), collected by an optical fiber (OF), measured by a photon counter (PT) and finally processed by a computer (PC).

The concentration of Coumarin102 solution is 100 μmol/L. It is prepared by anhydrous ethanol and deionized water. Phosphate buffer solution (PBS) is applied to adjust the electric conductivity of the two solutions to reach $\lambda$ =1:5400, 1:2000, 1:200 and 1:2 respectively.

Fig. 6(b) illustrates the schematic diagram of the EK flow microchip. The microchannel is 100 μm in high ($h$), 18 mm in long and 650 μm in width between two electrodes. The two streams with different electric conductivity are injected separately from inlets 1 and 2 by a syringe pump. Consequently, a pronounced electrical conductivity gradient is established immediately following the trailing edge.

The measurement point ($x_0$ in Fig. 6(c)) is 150 μm downstream of the trailing edge. $x_1$ serves as the velocity calibration location, which is about 2 mm downstream of the trailing edge. At this position, the pressure-driven basic flow is stable and fully developed. The velocity profile at the velocity calibration point is numerically simulated by COMSOL Multiphysics, as shown in Fig. 6(d), to reveal the relationship between fluorescent intensity and velocity[39]. A velocity calibration curve is plotted in Fig. 6(e). Fig. 6(f) shows a typical velocity time trace.

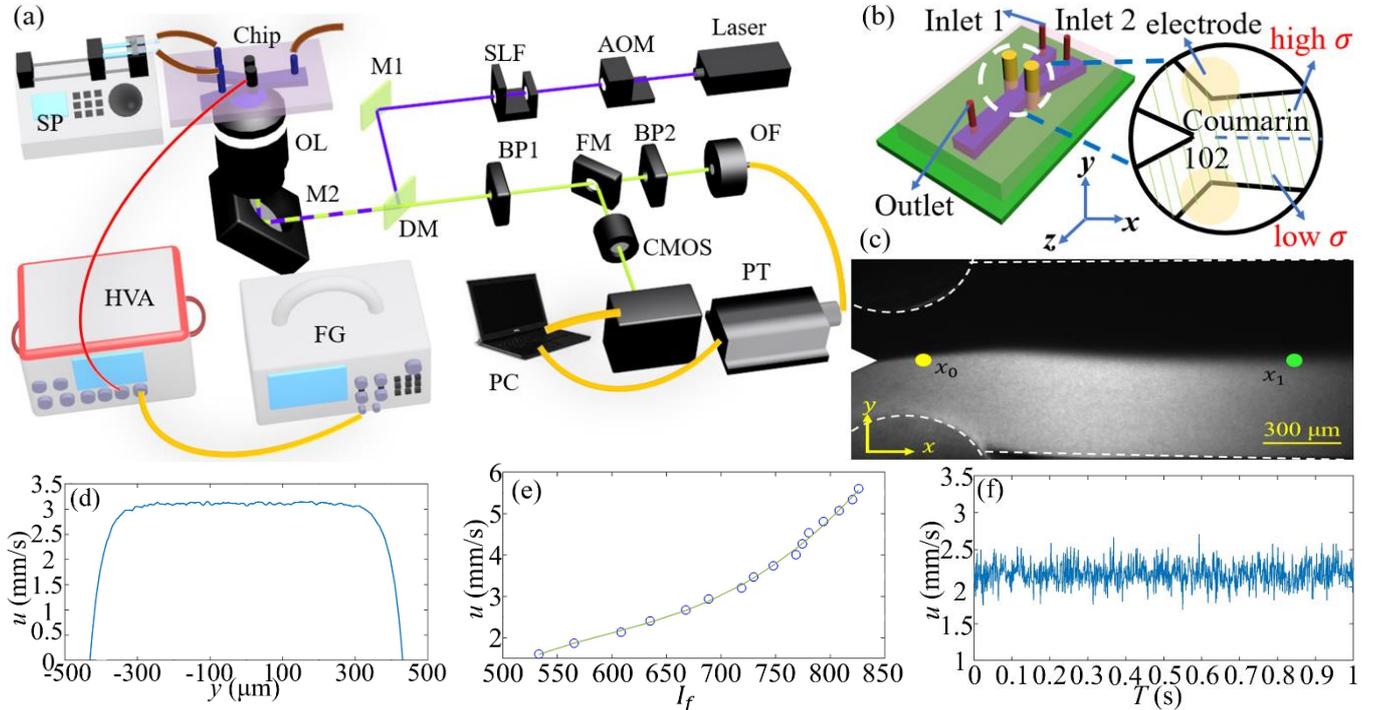

**FIG. 6.** Experimental system. (a) The diagram of LIFPA system. Laser: 405 nm continuous wave laser, AOM: Acousto-optic modulator, SLF: Spatial light filter, M1 and M2: mirrors, DM: Dichroic mirror, BP1 and BP2: Bandpass filters, FM: flipping mount mirror, OF: Photon counter, PT: Photon counter, PC: computer, FG: Function generator, HVA: High voltage amplifier, SP: Syringe pump, OL: Objective lens, CMOS: camera. (b) Structure diagram of the chip. (c) Flow visualization of the pressure-driven basic flow without electric field. $x_0$ is the measurement position. $x_1$ is the velocity calibration position. (d) The spanwise distribution of velocity in $x_1$ when $Q = 5$ μL/min. (e) The velocity calibration in the center of channel. (f) The velocity time trace of $\chi = 1.8 \times 10^{-4}$, $\lambda = 5400$ and $\Delta y = 11.4$ μm.

Through the analysis and comparison of the above two models, it is clear that the tanh model contains richer and more complex information about the influence of $\lambda$, $\chi$ and $\Delta y$ on $Ra_e$. To demonstrate the effectiveness of the tanh model, we studied the critical electric Rayleigh number ($Ra_{ec}$), when the EK flow becomes unstable, under different control parameters. $Ra_{ec}$ is determined from the power spectrum ($E(f)$) of the velocity fluctuations, as shown in Fig. 7(a). When no electric field was applied, a peak in $E(f)$, referred to as $E_{peak}$, is observed attributing to the perturbation of the wake after the trailing edge. When a smaller AC electric field was applied, i.e. $Ra_e < Ra_{ec}$, $E_{peak}$ is equivalent to that observed without electric field. While AC electric field is over some threshold value, $Ra_e > Ra_{ec}$, $E_{peak}$ increases rapidly with $Ra_e$, as plotted in Fig. 7(b), indicating an amplification of the flow perturbation induced by EBF. Through this way, $Ra_{ec}$ can be determined (Fig. 7(b)).

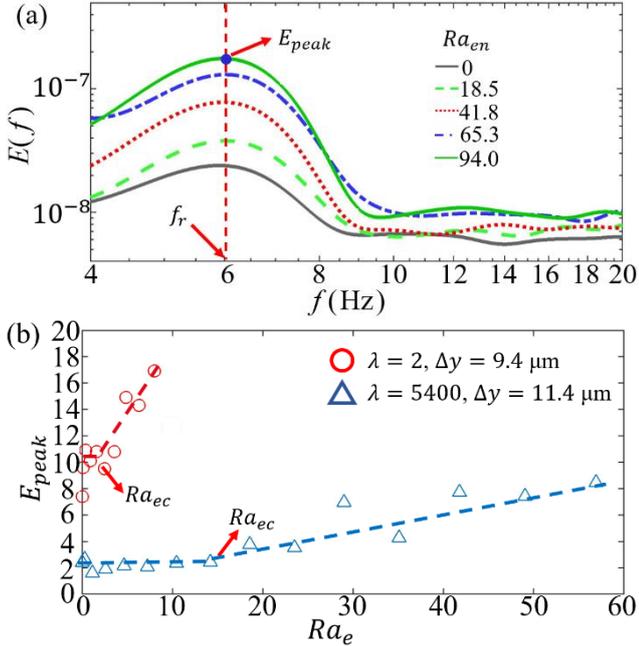

FIG. 7. Velocity power spectrum, $E_{peak}$ and the determination of $Ra_{ec}$. (a) Velocity spectrum $E(f)$ at $\chi = 1.1 \times 10^{-6}$, $\lambda = 5400$ and $\Delta y = 11.4$ μm. (b) $E_{peak}$ vs $Ra_e$ showing the position of $Ra_{ec}$ at $\chi = 1.1 \times 10^{-6}$.

## B. Experimental results

The relevant experimental (the symbols) and theoretical (the dotted lines) results on $Ra_{ec}$ are plotted in Figs. 8, 9 and 10 respectively. First, we analyze the influence of $\lambda$ on $Ra_{ec}$, as plotted in Fig. 8. When $\chi = 1.1 \times 10^{-6}$, $Ra_{ec}$ increases with $\lambda$ first before subsequently decreasing in both experimental and theoretical results. Although the experimental one exhibits stronger bending relative to the theoretical one, their variation shows consistent trending. The consistency between experimental and theoretical results increases as $\chi$ increases, e.g. at $\chi = 1.8 \times 10^{-5}$. When $\chi = 3.7 \times 10^{-3}$, $Ra_{ec}$ is monotonically increasing with $\lambda$. All the experimental trends align with the theoretical results predicted by the tanh model. In contrast, the most used model, i.e. linear distribution of $\sigma$, fails to provide even a qualitatively consistent prediction.

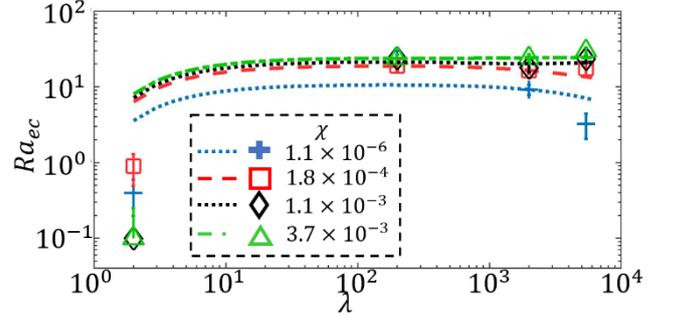

FIG. 8. $Ra_{ec}$ vs $\lambda$ under different $\chi$ with $\Delta y = 9.4$ μm.

Then, we further demonstrate the tanh model can effectively predict the influence of $\chi$ on $Ra_{ec}$. Both the experimental and theoretical results are plotted in Fig. 9. The experimental results indicate, when $\lambda \leq 200$, $Ra_{ec}$ is almost unchanged with $\chi$, which is consistent to the theoretical results. When $\lambda > 200$, the tanh model further reveals the complex variation of $Ra_{ec}$ with $\chi$, such that $Ra_{ec}$ keeps constant first, then increases until reaching a plateau. The results again support the effectiveness of the tanh model.

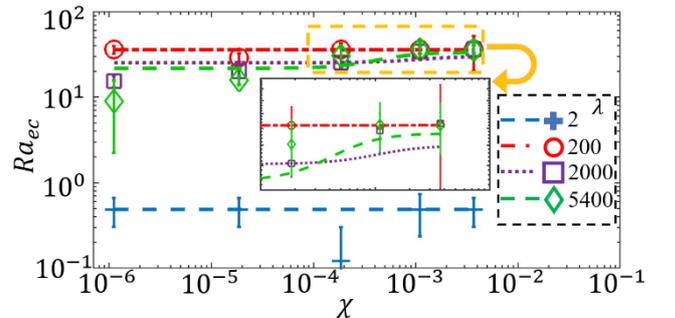

FIG. 9. $Ra_{ec}$ vs $\chi$ under different $\lambda$ with $\Delta y = 10.4$ μm.

The parameter $\Delta y$ is crucial to the gradient of electric conductivity, which in turn, determines EBF and the corresponding electric Rayleigh number. From Fig. 10(a), the experimental results of $Ra_{ec}$ are monotonically increasing with $\Delta y$, which is qualitatively consistent with the theoretical prediction by Eq. (15). Considering a larger $\Delta y$ is corresponding to a smaller electric conductivity gradient and smaller EBF, it indicates that the flow is more difficult to be disturbed at a larger $\Delta y$ under certain Reynolds number.

Finally, we also studied the influence of flow rate on $Ra_{ec}$, through bulk flow Reynolds number $Re = UL/\nu$, where $L = 2dh/(d+h)$ is the hydraulic diameter of the EK micromixers and $U$ is the bulk flow velocity. From Fig. 10(b), it can be seen that $Ra_{ec}$ has an inverse relationship with $Re$, indicating a decreasing requirement of AC electric field. Under a certain electric field, it is easier to generate unstable EK flow at a higher $Re$.

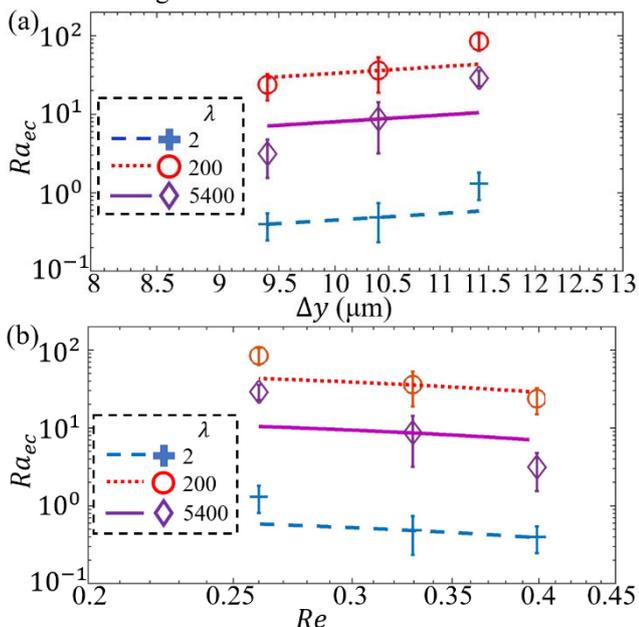

FIG. 10. (a) $Ra_{ec}$ vs $\Delta y$ under different $\lambda$ with $\chi = 1.1 \times 10^{-6}$. (b) $Ra_{ec}$ vs $Re$ under different $\lambda$ with $\chi = 1.1 \times 10^{-6}$.

## IV. Discussion

Dimensionless parameters are vital for engineering applications where complex relations among multiple control parameters are interplayed. An effective dimensionless parameter that provides insight into the underlying relationship among the control parameters can also enhance our understanding of the complex physical mechanisms in an engineering system. Unfortunately, when more control parameters are taken into account, it is significantly difficult to reach an effective dimensionless parameter. Part of the researchers believe that the pursuit of universal dimensionless parameter could be meaningless.

From this investigation, we aspire to show that, if a complex dimensionless parameter is appropriately expressed, it not only evaluates the status of the system, but also predicts (even though qualitatively) how the system changes. By using a more accurate distribution of electric conductivity, i.e. tanh model, the existence of optimal $\lambda$ and $\chi$ can be predicted through the electric Rayleigh number. The AC EK flow in a divergent microchannel can be unstable with $Ra_{ec} = O(10^{-1})$. In contrast, the critical electric Rayleigh numbers in previous reports, e.g. Chen et al[23] and Nan et al[21], are estimated to be $4.6 \times 10^3$ and 92.6 respectively, if using the tanh model in this investigation.

On the other hand, it should be noted that the current experimental and theoretical results, e.g. on the influence of $\Delta y$, still remain apparent discrepancies. This could be attributed to that the experimental environment is more complex relative to the theoretical model. $\Delta y$ of the electric conductivity interface varies along streamwise, making the electric conductivity distribution in experiments intrinsically 2D or more. To change $\Delta y$, we change the flow rates of the two streams, which leads to the change of $Re$ as well. It is almost impossible for us to isolate $\Delta y$ and $Re$ from each other in experiments. Thus, discrepancies become inevitable.

## V. Conclusion

In this investigation, we theoretically derive the expression of electric Rayleigh number $Ra_e$ for AC electrokinetic flow of binary fluids, on the basis of both linear and tanh distribution model of electric conductivity. The influence of electric conductivity ratio, AC frequency, interface width and electric field intensity on $Ra_e$ is systematically investigated in the tanh model, and compared with the linear model. Through comparison, it can be found that the tanh model provides more information about the variation of $Ra_e$ with the multiple parameters. In the tanh model, $Ra_e$ is more sensitive to electric conductivity ratio, AC frequency and interface width.

Furthermore, we conduct an experimental investigation on the critical electric Rayleigh number of the EK flow evaluated by the tanh model, through velocity measurement by LIFPA. The experimental findings exhibit an acceptable agreement with the theoretical predictions, thereby validating the efficacy of the electric Rayleigh number within the context of the tanh model.

An appropriate dimensionless parameter is not just a value, but an indicator of the underlying physical mechanism to guide engineering applications. In this paper, we propose a more accurate electric Rayleigh number to evaluate the state of AC EK flow, for designation of micromixers/reactors by electrokinetic approaches.

## ACKNOWLEGEMENTS

This research was funded by National Natural Science Foundation of China, grant number 51927804.